\documentclass[a4paper,11pt]{article}
\pdfoutput=1
\usepackage[english]{babel}
\usepackage{jheppub}
\usepackage{amsmath,amssymb,graphicx,textcomp,enumerate,xspace}
\usepackage{soul} 

\usepackage[utf8]{inputenc}

\newcommand\sss{\mathchoice%
{\displaystyle}%
{\scriptstyle}%
{\scriptscriptstyle}%
{\scriptscriptstyle}%
}

\def\beq{\begin{equation}}
\def\beqn{\begin{eqnarray}}
\def\eeq{\end{equation}}
\def\eeqn{\end{eqnarray}}

\def\lq{\left[} 
\def\rq{\right]} 
\def\rg{\right\}} 
\def\lg{\left\{} 
\def\({\left(} 
\def\){\right)}

\newcommand\as{\alpha_{\sss\rm S}}

\newcommand\CF{C_{\sss\rm F}}
\newcommand\CA{C_{\sss\rm A}}

\newcommand\TR{T_{\sss\rm R}}
\newcommand\NC{N_{\rm c}}

\newcommand\NF{n_{\rm f}}
\newcommand\muf{\mu_{\sss\rm F}}
\newcommand\mur{\mu_{\sss\rm R}}

\newcount\minutes 
\newcount\scratch 
\def\timestamp{%
\scratch=\time 
\divide\scratch by 60 
\edef\hours{\the\scratch} 
\multiply\scratch by 60 
\minutes=\time 
\advance\minutes by -\scratch 
---$\,$\hours:\null 
\ifnum\minutes< 10 0\fi 
\the\minutes}

\newcommand\qt{q_{\sss\rm T}}

\newcommand{\qtcut}{q_{\sss \rm T}^{\sss\rm cut}}
\newcommand{\qtmax}{q_{\sss \rm T}^{\sss\rm max}}
\newcommand{\zmax}{z^{\sss\rm max}}

\newcommand{\qtoQsq}{{\displaystyle\frac{\qt^2}{Q^2}}}

\newcommand\litw{{\rm Li}_2}
\newcommand\Litw{{\rm Li}_2}
\newcommand\lith{{\rm Li}_3}
\newcommand\Lith{{\rm Li}_3}

\def\beeq{\begin{eqnarray}} 
\def\eeeq{\end{eqnarray}} 
\def\to{\rightarrow}

\newcommand\HCCF{\lq \HCCF H^{F} C_1 C_2 \rq_{c\bar{c};\,ab}}

\newcommand\Lum{{\cal L}}
\newcommand\ord[1]{\mathcal{O}\!\(#1\)}

\newcommand\lumn{\Lum}
\newcommand\lumnn{\Lum^{'}}

\newcommand\APreg{p}

\newcommand\dsigdqtz{\frac{d\hat\sigma_{ab}\!\(\qt,z\)}{d\qt^2}}
\newcommand\dsigdqtzqg{\frac{d\hat\sigma_{qg}\!\(\qt,z\)}{d\qt^2}}

\newcommand\dontshow[1]{}

\newcommand\cgkzero{^{\sss c\!} g_{\sss 0}^{\sss H}}
\newcommand\gkzero{g_{\sss 0}^{\sss H}}
\newcommand\cgkone{^{\sss c\!} g_{\sss 1}^{\sss H}}
\newcommand\cgktwo{^{\sss c\!} g_{\sss 2}^{\sss H}}
\newcommand\gkone{g_{\sss 1}^{\sss H}}
\newcommand\gktwo{g_{\sss 2}^{\sss H}}

\newcommand\cJk{^{\sss c\!}{\cal J}^{\sss H}_{\sss 23}}
\newcommand\Jk{{\cal J}^{\sss H}_{\sss 23}}

\title{Power corrections in a transverse-momentum cut for 
  vector-boson production at NNLO: the $\boldsymbol{qg}$-initiated real-virtual contribution}

\author[a,b]{Carlo Oleari,}
\author[a,b]{Marco Rocco}

\emailAdd{carlo.oleari@mib.infn.it}
\emailAdd{m.rocco10@campus.unimib.it}

\affiliation[a] {INFN, Sezione di Milano\,-\,Bicocca,
  Piazza della Scienza 3, 20126 Milano, Italy}

\affiliation[b] {Universit\`a di Milano\,-\,Bicocca,
  Piazza della Scienza 3, 20126 Milano, Italy}

\abstract{We consider the production of a vector boson ($Z$, $W^\pm$ or
  $\gamma^*$) at next-to-next-to-leading order in the strong coupling
  constant $\as$.
  We impose a transverse-momentum cutoff, $\qtcut$, on the vector boson
  produced in the $qg$-initiated channel. We then compute the power
  corrections in the cutoff, up to the second power, of the real-virtual
  interference contribution to the cumulative cross section at order $\as^2$.
  Other terms with the same kinematics, originating from the subtraction
  method applied to the double-real contribution, have been also considered.
  
  The knowledge of such power corrections is a required
  ingredient in order to reduce the dependence on the transverse-momentum
  cutoff of the QCD cross sections at next-to-next-to-leading order, when the
  $\qt$-subtraction method is applied. In addition, the study of the
  dependence of the cross section on $\qtcut$ allows as well for an
  understanding of its behaviour in the small transverse-momentum limit,
  giving hints on the structure at all orders in $\as$ and on the
  identification of universal patterns.

  Our result are presented in an analytic form, using the process-independent
  procedure described in a previous paper for the calculation of the
  all-order power corrections in $\qtcut$.
  
}


\begin{document}

\maketitle

\section{Introduction}
\label{sec:intro}

The recent years have witnessed an increasing growth in the accuracy of
physics measurements at the Large Hadron Collider, on the one side, and the
great efforts done by the theoretical community in order to provide
theoretical results of increasing precision, on the other. The goal of these
activities is not only important for the extraction of Standard Model~(SM)
parameters, but also for searches of signals of new physics that can appear
as small deviations with respect to the SM predictions.

Reaching the highest possible level of precision is then the main goal and
the calculation of perturbative QCD corrections plays a dominant role in this
context.  Until a few years ago, the standard for such calculations was
next-to-leading-order~(NLO) accuracy. In recent years, the goal has become
next-to-next-to-leading-order~(NNLO) accuracy, and even beyond for
some processes.

The computation of higher-order terms becomes more involved due to the
technical difficulties arising in the evaluation of virtual contributions and
to the increasing complexity of the infrared~(IR) structure of the real
contributions. In order to expose the cancellation of the IR divergences
between real and virtual contributions, the knowledge of the behaviour of the
scattering amplitudes in the infrared limits is then crucial and it is indeed
what is used by the subtraction methods in order to work.

At NNLO and beyond, several subtraction schemes have been proposed in the
past years. These schemes mostly fit into two categories: local methods and
slicing methods. The latter are based on partitions of the phase space into
hard regions and infrared-sensitive regions, where the cancellation of
divergences is performed with non-local subtraction terms.  In order to
apply these methods, one has to introduce a resolution parameter to
identify the phase-space regions where the non-local subtraction
acts. Slicing methods that have been successfully applied at NNLO and N$^3$LO
are the transverse-momentum~($\qt$) subtraction method~\cite{Catani:2007vq,
  Bozzi:2005wk, Bonciani:2015sha, Cieri:2018oms, Catani:2019iny} and
$N$-jettiness subtraction~\cite{Boughezal:2015eha, Gaunt:2015pea}.

By applying non-local subtraction methods, the singular terms in the
small-cutoff limit are cancelled. These terms have a universal nature and
this allows to construct the subtraction terms on general grounds.  After the
cancellation has taken place, only finite and vanishing terms remain. These
terms are, in general, process dependent.  A residual dependence on the
cutoff then remains as power corrections. While these terms formally vanish
in the null-cutoff limit, they give a non-zero numerical contribution for any
finite choice of the cutoff.

The knowledge of the power-correction terms greatly increases the numerical
reliability of the final results. In fact, by subtracting the lowest powers
in the cutoff makes the result less sensitive to the arbitrary cutoff,
numerically approaching the theoretical limit of this parameter going to
zero.  This is not only valid when the subtraction method is applied to NLO
computations, but it is numerically more relevant when applied to
higher-order calculations, as pointed out, for example, in the evaluation of
NNLO cross sections in refs.~\cite{Grazzini:2017mhc, Boughezal:2016wmq}.

Beyond reducing the dependence of the theoretical results on the cutoff, the
study of power-suppressed terms in the infrared regions is a theoretically
interesting subject, since it allows to deepen our knowledge of the universal
and non-universal structure of the perturbative behaviour of QCD cross
sections in the IR limits. Thus, several papers have tackled the study of
power corrections in the general framework of fixed-order and
threshold-resummed computations~\cite{vanBeekveld:2019cks,
  vanBeekveld:2019prq, DelDuca:2017twk, Bonocore:2015esa, Bonocore:2014wua,
  Laenen:2010uz, Laenen:2008ux, Beneke:2018gvs}.

Power corrections at NLO have been extensively studied in
refs.~\cite{Moult:2016fqy, Boughezal:2016zws, Boughezal:2018mvf,
  Moult:2017jsg, Ebert:2018lzn, Bhattacharya:2018vph, Campbell:2019gmd,
  Moult:2018jjd, Boughezal:2019ggi, Ebert:2019zkb, Ebert:2020dfc} in the
context of the $N$-jettiness subtraction method, and in
refs.~\cite{Bauer:2000ew, Bauer:2000yr, Bauer:2001ct, Bauer:2001yt,
  Bauer:2002aj, Moult:2019mog} within SCET-based subtraction methods. Power
corrections at NLO for the transverse momentum of a colour singlet have been
derived for the first time at differential level in ref.~\cite{Ebert:2018gsn}
within the SCET framework.
In ref.~\cite{Cieri:2019tfv}, we presented a method to compute the power
corrections at all orders, for the inclusive production of a colourless
final-state system, at NLO in QCD.  Recently, the leading power corrections
for the electroweak NLO corrections to the inclusive cross section for the
production of a massive lepton pair through the Drell--Yan mechanism have been
computed in ref.~\cite{Buonocore:2019puv}.

$N$-jettiness power corrections at NNLO have been considered in
refs.~\cite{Moult:2016fqy, Moult:2017jsg}. In particular, analytic results
are obtained for the dominant $\as\,\tau\log(\tau)$ and
$\as^2\,\tau\log^3(\tau)$ subleading terms, where $\tau$ is the 0-jettiness,
for $q\bar{q}$-initiated Drell--Yan production and for $gg$-, $gq$- and
$q\bar{q}$-initiated Higgs boson production, along with a numerical fit for
the subdominant terms.  

In this paper we consider the production of a vector boson ($Z$, $W^\pm$ or
$\gamma^*$) at NNLO in the strong coupling constant $\as$.
We impose a transverse-momentum cutoff, $\qtcut$, on the vector boson
produced in the $qg$-initiated channel, and we compute, for the first time,
the power corrections in the cutoff, up to the second power $(\qtcut)^2$, of
the real-virtual interference contribution to the cumulative cross section at
order $\as^2$, plus other terms with the same kinematics, originating from
the application of the subtraction method to the double-real contribution.
In order to perform this computation, we apply the general
process-independent method that we have formulated in
ref.~\cite{Cieri:2019tfv}.

This is the first step in order to compute the power corrections, up to the
second power, of the NNLO cumulative cross section for vector-boson
production.  In fact, together with the real-virtual $qg$-initiated channel
that we consider in this paper, also the $qq$-initiated channel
contributes to the real-virtual terms, together with all the double-real
radiation contributions. We will consider these contributions in future
works.

The outline of this paper is as follows. In Sec.~\ref{sec:kinematics} we
introduce our notation and we briefly summarize the expressions of the
partonic and hadronic cross sections, in a form that is suitable for what
follows. In Sec.~\ref{sec:calculation} we outline the calculation we have
done and in Sec.~\ref{sec:results} we present and discuss our analytic
results. We draw our conclusions in Sec.~\ref{sec:conclusions}.  In
Appendix~\ref{app:sample_z-int}, we present some examples of the integrals we
had to perform in order to compute the power corrections, and we give the
result of the integration for a few of them. Finally, in
Appendix~\ref{app:rvFqg} we collect the final results of our paper.

\section{The hadronic and partonic cross sections}
\label{sec:kinematics}
In this section we set the theoretical framework and introduce the notation
used throughout the paper.

\subsection{The hadronic cross sections}
We consider the production of a colourless system $F$ with quadri-momentum
$q$ and squared invariant mass $Q^2$, plus a coloured system $X$ at a hadron
collider
\begin{equation}
\label{eq:proc}
h_1+h_2 \to F+X \,.
\end{equation}
We call $S$ the hadronic squared center-of-mass energy and we write the hadronic
differential cross section for this process as
\begin{equation}
\label{eq:had_tot_XS}
d\sigma = \sum_{a,b} \int_\tau^1 dx_1 \int_\frac{\tau}{x_1}^1 dx_2\,
f_a\!\(x_1\) f_b\!\(x_2\) d\hat\sigma_{ab}\,,
\end{equation}
where 
\begin{equation}
\label{eq:tau_def}
\tau = \frac{Q^2}{S}\,,
\end{equation}
$f_{a/b}$ are the parton densities of the partons $a$ and $b$, in the hadron
$h_1$ and $h_2$ respectively, and $d\hat\sigma_{ab}$ is the partonic cross
section for the process $a+b \to F + X$. The dependence on the
renormalisation and factorisation scales and on the other kinematic
invariants of the process are implicitly assumed. 

The hadronic cross section can be written as\footnote{For more details, see
Appendix~A of ref.~\cite{Cieri:2019tfv}.}
\begin{equation}
\sigma = \sum_{a,b} \int_\tau^1 dx_1 \int_\frac{\tau}{x_1}^1 dx_2\,
f_a\!\(x_1\) f_b\!\(x_2\) \int  d\qt^2\, dz \,\dsigdqtz\,
\delta\!\(\! z- \frac{Q^2}{s}\),
\end{equation}
where $s$ is the partonic center-of-mass energy, equal to
\begin{equation}
\label{eq:s_S}
s = S \, x_1\, x_2\,.
\end{equation}
We have also made explicit the dependence on $z$, the ratio between the
squared invariant mass of the system $F$ and the partonic center-of-mass
energy, and on $\qt$, the transverse momentum of the system $F$ with respect
to the hadronic beams.  Using eqs.~(\ref{eq:s_S}) and~(\ref{eq:tau_def}) and
integrating over $x_2$ we obtain
\begin{equation}
\sigma = \sum_{a,b} \tau \int_\tau^1 \frac{dz}{z}  \int_\frac{\tau}{z}^1 \frac{dx_1}{x_1}\,
f_a\!\(x_1\) f_b\!\(\frac{\tau}{z\,x_1}\) \frac{1}{z} \int\! d\qt^2 \, \dsigdqtz \,.
\end{equation}
We then introduce the parton luminosity ${\cal L}_{ab}(y)$ defined by
\begin{equation}
  \label{eq:lum_def}
{\cal L}_{ab}(y) \equiv \int_y^1 \frac{dx}{x}\, f_a\!\(x\) f_b\! \(\frac{y}{x}\),
\end{equation}
so that we can finally write
\begin{equation}
\label{eq:sigma_had_z_one}
\sigma = \sum_{a,b} \tau \int_\tau^1 \frac{dz}{z}  \, {\cal L}_{ab}\!\(\frac{\tau}{z}\)
 \frac{1}{z} \int\! d\qt^2 \,  \dsigdqtz \,.
\end{equation}

\subsection{The partonic differential cross sections}
\label{sec:part_diff_XS}
In this paper we consider the NNLO corrections to the production of a vector
boson $F$, i.e.~a $W^\pm$, a $Z$ or a virtual photon $\gamma^*$.
In particular, we deal with the $qg$-initiated partonic channel
\begin{equation}
  \label{eq:qg_VX}
    q(p_1) + g(p_2) \,\,\rightarrow\,\, F(q) + X(k) \,,
\end{equation}
where the quadri-momenta are given in parentheses.  In
ref.~\cite{Cieri:2019tfv}, among other contributions, we considered the NLO
cross section for $F$ production, i.e.
\begin{equation}
  \label{eq:qg_Vq}
    q(p_1) + g(p_2) \,\,\rightarrow\,\, F(q) + q(k) \,,
\end{equation}
where the final-state quark has the same flavour of the initial-state one, for
$Z/\gamma^*$ production, and different flavour, for $W$ production. We
generically indicate the initial- and final-state quark with the same letter $q$.

Introducing the kinematic invariants
\begin{equation}
  s = (p_1 + p_2)^2\,, \qquad  t = (p_1-q)^2\,, \qquad   u = (p_2-q)^2, 
\end{equation}
we have the relation
\begin{equation}
  s + t + u = q^2 + s_2
\end{equation}
where $s_2 = k^2$ is the squared invariant mass of the system recoiling
against the $F$ boson at parton level.

In the following, we use the same notation and the expressions computed in
ref.~\cite{Gonsalves:1989ar}. The couplings appearing in the differential
cross sections follow this convention: if an electroweak boson $F$ is emitted
by a quark with flavour $f_1 = \{u,\,d,\,s,\,c,\,b\}$ which then
changes into $f_2$, the vertex is described by the Feynman rule
\begin{equation}
  -i e \gamma^\mu \lq \ell_{\sss f_2 f_1} \, \frac{1-\gamma_5}{2} +
  r_{\sss f_2 f_1} \, \frac{1+\gamma_5}{2} \rq ,
\end{equation}
where the definitions of the left- and right-handed couplings $\ell$ and $r$
depend on the $F$ boson
\begin{eqnarray}  
  W^{-} \,&:& \,\,\, \ell_{\sss f_2 f_1} = \frac{1}{\sqrt{2}\,\sin\theta_{\sss W}} \,
  (\sigma_{\sss +})_{\sss f_2 f_1} \, V_{\sss f_2 f_1} \,,   \qquad r_{\sss f_2 f_1} = 0 \,,
  \\
  W^{+} \,&:& \,\,\, \ell_{\sss f_2 f_1} = \frac{1}{\sqrt{2}\,\sin\theta_{\sss W}} \,
  (\sigma_{\sss -})_{\sss f_2 f_1} \, V_{\sss f_2 f_1}^\dagger \,,   \qquad r_{\sss f_2 f_1} = 0 \,,
  \\
  Z \,&:& \,\,\, \ell_{\sss f_2 f_1} = \frac{1}{\sin 2\theta_{\sss W}} \,
  (\sigma_{\sss 3})_{\sss f_2 f_2} - \delta_{\sss f_2 f_1} \, e_{f_1} \tan\theta_{\sss W} \,,
  \qquad r_{\sss f_2 f_1} = - \delta_{\sss f_2 f_1} \, e_{\sss f_1} \tan\theta_{\sss W} \,,
  \phantom{aaaaa}
  \\[2mm]
  \gamma^* \,&:& \,\,\, \ell_{\sss f_2 f_1} = r_{\sss f_2 f_1} =
  \delta_{\sss f_2 f_1} \, e_{f_1} \,,
\end{eqnarray}
where $\theta_{\sss W}$ is the Weinberg angle, $e_{\sss f}$ is the fractional
electric charge of the quark with flavour $f$, $\sigma_{\sss \pm} =
(\sigma_{\sss 1}\pm i\sigma_{\sss 2})/2$ and $\sigma_{\sss 3}$ are the weak
isospin Pauli matrices and $V$ is the unitary Cabibbo–Kobayashi–Maskawa
mixing matrix. In addition, in the following we abbreviate $\ell_{\sss f_2 f_1}$  to
$\ell_{\sss 21}$, and the same  for $r_{\sss f_2 f_1}$.

The QCD NLO corrections to eq.~(\ref{eq:qg_VX}) were
computed in ref.~\cite{Gonsalves:1989ar}. We report here eq.~(2.12) of this
reference, since we are going to use their results  in $d=4$
space-time dimensions, after correcting for some known typos\footnote{See
footnote $\S$ of ref.~\cite{Catani:2012qa}.} 
\begin{eqnarray}
  \label{eq:gonsalves_Vqg}
  E_q \, \frac{d\hat{\sigma}_{ qg}}{d^3 q} &=&
  \frac{1}{s}  \,\frac{\CF}{\NC^2-1} \, \alpha\, \as(\mur)
  \Bigg\{ \delta(s_2) \, A^{qg}\!\(s,t,u\)
  \sum_f \(\left| \ell_{\sss f1}\right|^2 + \left| r_{\sss f1}\right|^2 \)
  \nonumber\\
  &&{} + \frac{\as(\mur)}{2 \pi} \bigg\{ \bigg[
    \delta(s_2) \Big( B_1^{qg}\!\(s,t,u\) + \NF \, B_2^{qg}\!\(s,t,u\) +
      C_1^{qg}\!\(s,t,u\) + C_2^{qg}\!\(s,t,u\) \Big)
  \nonumber\\
  && \hspace{2.2cm} {}  +  C_3^{qg}\!\(s,t,u,s_2\) \bigg]
  \sum_f \(\left| \ell_{\sss f1}\right|^2 + \left| r_{\sss f1}\right|^2 \)
   \nonumber\\
   && \hspace{1.9cm}  
  {}+ \delta(s_2) \, B_3^{qg}\!\(s,t,u\)  \(\ell_{\sss 11}-r_{\sss 11}\)
  \sum_f \(\ell_{ \sss f f}-r_{\sss ff}\)  \bigg\} \Bigg\} \,,
\end{eqnarray}
where $E_q$ is the energy of the $F$ boson, $\NC=3$ is the number of colours
and $\CF=(\NC^2-1)/(2\NC) =4/3$.  The functions $A^{qg}$, $B^{qg}_i$,
$C^{qg}_i$ ($i=1,2,3$) are defined in eqs.~(A4)--(A6),~(A10)--(A12) of
ref.~\cite{Gonsalves:1989ar}. $A^{qg}$ is the contribution at tree level of
the process in eq.~(\ref{eq:qg_Vq}). The functions $B_i^{qg}$ receive
contributions from the interference of the one-loop virtual corrections to
eq.~(\ref{eq:qg_Vq}), with the tree-level contribution. In particular,
$B_2^{qg}$ originates from the renormalisation counterterm, while $B_3^{qg}$ is
the contribution from the virtual diagrams with a triangular quark loop,
which are present only for $Z/\gamma^*$ production.  These contributions are
then multiplied by a $\delta(s_2)$ term, since the system recoiling against
the $F$ boson only comprises a single quark with momentum $k$, so that
$s_2=k^2=0$. 

The functions $C_i^{qg}$ originate from the diagrams contributing to the real
corrections. In particular, $C_1^{qg}$ and $C_2^{qg}$ are the coefficient of
a $\delta(s_2)$ term, leftovers of the subtraction method when dealing with
initial- and final-state radiation. $C_3^{qg}$ contributes instead for
non-zero values of $s_2$, and corresponds to the double-real radiation
contribution to $qg$-initiated $F$ boson production.  In the following we
neglect all the infrared divergences appearing as poles in
eqs.~(A4)--(A6),~(A10)--(A12) of ref.~\cite{Gonsalves:1989ar}, since they
cancel out when summing real and virtual contributions at this order in $\as$.

The $B_i^{qg}$ and $C_i^{qg}$ are analytic functions of the kinematic
invariants and contain logarithmic and dilogarithmic functions.

In this paper we present results for the calculation of the power corrections
for all the terms proportional to $\delta(s_2)$ in
eq.~(\ref{eq:gonsalves_Vqg}), i.e.~the virtual-correction terms and terms
from the regularisation of the double-real radiation contributions.

Since the kinematics of these terms is equivalent to the one discussed in
ref.~\cite{Cieri:2019tfv}, we follow the same procedure described in its
Appendix~A (in particular eqs.~(A.16)--(A.20)), and we integrate all the
terms proportional to $\delta(s_2)$ in eq.~(\ref{eq:gonsalves_Vqg}), writing
them in the form suitable to be inserted in eq.~(\ref{eq:sigma_had_z_one}),
i.e.
\begin{equation}
\label{eq:diffXS_M}
\left.\dsigdqtzqg \right|_{\delta(s_2)} = \frac{1}{16\pi} \frac{z^2}{Q^4}
\frac{1}{\sqrt{(1-z)^{2}-4z\,\qtoQsq}}
\lq \left| {\cal M}\(z,t_+,\qt\)\right|^2  + \left| {\cal
  M}\(z,t_-,\qt\)\right|^2\rq ,
\end{equation}
where ${\cal M}(s,t,u)$ is the sum of the functions $A^{qg},
B_1^{qg},B_2^{qg},B_3^{qg}, C_1^{qg}, C_2^{qg}$, as they appear in
eq.~(\ref{eq:gonsalves_Vqg}), together with the global factor in front,
evaluated at
\begin{equation} 
u=Q^2-s -t\,, \qquad   s=\frac{Q^2}{z}\,, \qquad t=t_\pm \,, 
\end{equation}
where
\begin{equation}
\label{eq:tpm_z}
t_\pm = \frac{Q^{2}}{2z}\,\lq z-1 \pm \sqrt{(1-z)^{2}-4z\,\qtoQsq} \rq ,
\end{equation}
so that ${\cal M}$ becomes a function of $z$ and $\qt$, for a given
vector-boson virtuality $Q^2$.

We can write eq.~(\ref{eq:gonsalves_Vqg}), manipulated according to the
previous steps, in a compact notation as
\begin{equation}
  \dsigdqtzqg = {\frac{\as}{2\pi}} \frac{d\hat\sigma_{qg}^{\sss (1)}(\qt,z)}{d\qt^2}
  + {\(\frac{\as}{2\pi}\)^2} \, \frac{d\hat\sigma_{qg}^{\sss (2)}(\qt,z)}{d\qt^2} ,
\end{equation}
where the superscript $^{(1)}$ denotes the tree-level cross section, while
the superscript $^{(2)}$ the virtual and real contributions.  The choice is
made in order to make contact with the labeling of the transverse-momentum
resummation coefficients, that refer to $F$ production as the zeroth term, to
its NLO corrections as the first term, and to the NNLO corrections, i.e. the
QCD NLO corrections to $F$ + 1 parton, as the second one.

In the rest of the paper we focus on the contribution 
\begin{equation}
\label{eq:dsig2}
 \left. \frac{d\hat\sigma_{qg}^{\sss (2)}(\qt,z)}{d\qt^2} \right|_{\delta(s_2)}  
\end{equation}
and, with a little abuse of notation, when referring to eq.~(\ref{eq:dsig2}),
we sometimes drop the~$\big|_{\delta(s_2)}$, to ease the notation.

\section{Description of the calculation}
\label{sec:calculation}
In order to compute the power corrections of the cross section in
eq.~(\ref{eq:dsig2}), we follow the path along which we proceeded in
ref.~\cite{Cieri:2019tfv} and which is described in Sec.~3 therein.

We recall here that, in the phase-space region where $\qt$ is different from
zero and much smaller than the invariant mass of the colour singlet, the
cross section of eq.~(\ref{eq:dsig2}) is characterised by a well-known
perturbative structure. In fact, it contains logarithmically-enhanced terms
that are singular in the $\qt\to 0$ limit~\cite{Dokshitzer:1978yd,
  Dokshitzer:1978hw, Parisi:1979se, Curci:1979bg, Collins:1981uk,
  Kodaira:1981nh, Kodaira:1982az, Collins:1984kg, Catani:1988vd,
  deFlorian:2000pr}, terms that are finite in the same limit, and power terms
that vanish in the small-$\qt$ limit.

It is customary in the literature~\cite{deFlorian:2001zd, Ebert:2018lzn} to
compute the following cumulative partonic cross section, integrating the
differential cross section in the range $0 \le \qt \le \qtcut$, in order to
derive the perturbative behaviour of these terms at small $\qt$
\begin{equation}
  \label{eq:XS_cumul}
 \hat\sigma_{ab}^{\sss  <}(z) \equiv
\int_0^{\(\qtcut\)^2}\!\! d\qt^2 \,\frac{d\hat\sigma_{ab}(\qt,z)}{d\qt^2} \,.
\end{equation}
For $F$ + 1 parton production at NLO, eq.~(\ref{eq:XS_cumul}) receives
contributions from the Born diagrams, that were analysed in
ref.~\cite{Cieri:2019tfv}, and from the virtual and real QCD corrections. The
former are proportional to $\delta(s_2)$, while the latter describe the
production of a further parton. Since the total NNLO partonic cross section for
$F$ production is finite, following what was done in
refs.~{\cite{Catani:2011kr, Catani:2012qa}, we compute the above integral as
  a difference
\begin{equation}
  \label{eq:int_0^qtcut-sub}
 \hat\sigma_{ab}^{\sss <}(z)  = \hat\sigma^{\rm\sss tot}_{ab}(z) -
 \hat\sigma_{ab}^{\sss >}(z)\,, 
\end{equation}
with
\begin{equation}
   \label{eq:int_qtcut^qmax}
  \hat\sigma^{\rm\sss tot}_{ab}(z)  =
 \int_0^{\(\qtmax\)^2}\!\! d\qt^2 \,\frac{d\hat\sigma_{ab}(\qt,z)}{d\qt^2}\,,
\qquad \qquad
\hat\sigma_{ab}^{\sss >}(z) = \int_{\(\qtcut\)^2}^{\(\qtmax\)^2}\!\!
d\qt^2\,\frac{d\hat\sigma_{ab}(\qt,z)}{d\qt^2} \,,
\end{equation}
where $\qtmax$ is the maximum transverse momentum allowed by the kinematics,
$\hat\sigma^{\rm\sss tot }_{ab}(z)$ is the total partonic cross section, and
$\hat\sigma_{ab}^{\sss >}(z)$ is the partonic cross section integrated above
$\qtcut$, that can be then computed in four space-time dimensions.

In this paper we study the $\qtcut \ll Q$ behaviour of the NNLO real-virtual
contribution to $F$ production, by computing the $\qtcut$-expansion of
\begin{equation}
\label{eq:dXS0dqt_V_gq}
\hat\sigma_{qg}^{\sss > (2)}(z)
=  \int_{\(\qtcut\)^2}^{\(\qtmax\)^2}\!\!d\qt^2 \,
\left. \frac{d\hat\sigma_{qg}^{\sss (2)}(\qt,z)}{d\qt^2}  \right|_{\delta(s_2)}  ,
\end{equation} 
up to $\ord{(\qtcut)^2}$ included. The integration goes from an arbitrary
value $\qtcut$ up to the maximum transverse momentum $\qtmax$ allowed by the
kinematics of the event, given by
\begin{equation}
\(\qtmax\)^2 = Q^2\, \frac{(1-z)^2}{4\,z}\,,
\end{equation}
at a fixed value of $z$.

The integration in $\qt$ is performed with dedicated changes of variables, in
order to get rid of the square roots within the arguments of logarithms and
dilogarithmic functions. The correct analytic continuation is then performed
in order to obtain a real result.
At difference with what was done in ref.~\cite{Cieri:2019tfv}, we do not
quote here the results of the integration, due to their length.

To lighten up the notation, we introduce the dimensionless
quantity\footnote{In the literature, the parameter $a$ is also referred to as
  $r^2_{\rm \sss cut}$ (see e.g.~\cite{Grazzini:2017mhc}).}
\begin{equation}
  \label{eq:a_def}
a \equiv  \frac{\(\qtcut\)^2}{Q^2}\,,
\end{equation}
that will be the expansion parameter in the rest of the paper.
Then, in order to compute the hadronic cross section of
eq.~(\ref{eq:sigma_had_z_one}), we need to integrate the partonic cross
sections convoluted with the corresponding luminosities.  In the calculation
of the total cross sections, the upper limit in the $z$ integration is
unrestricted and equal to 1. When a cut on the transverse momentum $\qt$ is
applied the $z$-integration range is instead bound from above, i.e.
\begin{equation}
\label{eq:def_fa}  
0 \le z \le \zmax \equiv 1-f(a)\,, \qquad \qquad f(a) \equiv 2
\sqrt{a}\(\sqrt{1 + a} - \sqrt{a}\).
\end{equation}
However, in order to make contact with the transverse-momentum subtraction
formulae, which allows us to recover the logarithmic-enhanced behaviour along
with the power corrections, we need to extend the integration range of the
$z$ variable up to 1 and then expand our results in powers of $a$. To this
aim, we used the same procedure presented in ref.~\cite{Cieri:2019tfv} in
order to deal with the divergent terms in the $z \to 1$ limit. The procedure
is very technical and all the details are presented in Appendix~B of the same
reference. Hence we refer the interested reader to that appendix for the
description of the method.

\section{Results}
\label{sec:results}
In this section we collect fully-analytic results for the NNLO power
corrections in the transverse-momentum cutoff, up to order $a$. The results
refer to the $\delta(s_2)$ contributions of the $qg$-initiated channel
in eq.~(\ref{eq:gonsalves_Vqg}).

We label the different contributions of eq.~(\ref{eq:gonsalves_Vqg}) with the
letter $K$, so that
\begin{equation}
  K= \lg A^{qg},\, B_1^{qg},\,B_2^{qg},\,B_3^{qg},\,C_1^{qg},\,C_2^{qg} \rg.
\end{equation}
Using eq.~(\ref{eq:diffXS_M}) and following the discussion in
Sec.~\ref{sec:part_diff_XS}, we integrate $K(s,t,u)$ in $t$ to obtain a
function of  the transverse momentum of the vector boson, $\qt$, and
$z$
\begin{equation}
  \int dt \, K(s,t,u) = K(\qt,z) \,.
\end{equation}
As recalled in Sec.~\ref{sec:calculation}, the functions $K$ are then
integrated in $\qt$ from an arbitrary value, $\qtcut$, up to the maximum
transverse momentum $\qtmax$ allowed by the kinematics of the event. 

We further split the contributions of $B_1^{qg}$ and $C_1^{qg}$ according to
their colour factor, $\CA$ and $\CF$. We then introduce a further index,
$c=\lg\CA,\CF\rg$, relevant for $B_1^{qg}$ and $C_1^{qg}$, in order to
distinguish the coefficients of the different colour factors.

The general procedure described in Appendix~B of ref.~\cite{Cieri:2019tfv} is
applied to the $\qt$-integrated $K$ functions. In order to present the
structure of the results, we refer to the definitions of $I$, $\tilde{I}_{1}$,
$\tilde{I}_{2}$ and $\tilde{I}_{3}$ of eqs.~(B.8)--(B.11) in
ref.~\cite{Cieri:2019tfv}.
Moreover, we present the results for the sum $\tilde{I}_{23} \equiv
\tilde{I}_{2}+\tilde{I}_{3}$, and we do not give the two terms
separately.

%
After dropping the $qg$ superscript for ease of notation, we can then write
\begin{equation}
\label{eq:I_final_nnlo}
I^{\sss H} = \tilde{I}_{\sss 1}^{\sss H} +\tilde{I}_{\sss 23}^{\sss H} \,, \qquad\qquad
H = \lg A, \, B_1,\,B_2,\,B_3,\,C_1,\,C_2 \rg \,,
\end{equation}
where, if $H = \lg B_1,\,C_1 \rg$,
\begin{eqnarray}
\label{eq:Itilde1_nnlo1}
\tilde{I}_{\sss 1}^{\sss H} &=& {} \sum_{c\,=\lg\CA,\,\CF\rg} c \lg  \int_0^1 d z\,
l(z) \; \cgkzero(z)
+  \int_0^1 d z \, l(z) \!\lq \frac{\cgkone(z)}{1-z} \rq_+
+  \int_0^1 d z \, l(z) \!\lq \frac{\cgktwo(z)}{1-z} \rq_{++} \rg ,
\nonumber
\\
 \\[2mm]
  \tilde{I}_{\sss 23}^{\sss H} &=& {} \sum_{c\,=\lg\CA,\,\CF\rg} c\,\, \cJk \,,  
\end{eqnarray}
while, if $H=\lg A,\,  B_2,\, B_3,\, C_2 \rg$,
\begin{eqnarray}
\tilde{I}_{\sss 1}^{\sss H} &=& {} \int_0^1 d z\, l(z) \; \gkzero(z)
+  \int_0^1 d z \, l(z) \!\lq \frac{\gkone(z)}{1-z} \rq_+
+  \int_0^1 d z \, l(z) \!\lq \frac{\gktwo(z)}{1-z} \rq_{++},
  \\[2mm]
  \label{eq:Itilde23_nnlo2}
\tilde{I}_{\sss 23}^{\sss H} &=& {} \Jk\,,
\end{eqnarray}
where $l(z)$ is given in terms of the parton luminosity in
eq.~(\ref{eq:lum_def}), where we have dropped any subscript for ease of
notation
\begin{equation}
  \label{eq:def_lz}
  l(z) \equiv \frac{1}{z}\,{\cal L}\(\frac{\tau}{z}\) \,.
\end{equation}
The functions $\cgkzero(z)$, $\cgkone(z)$, $\cgktwo(z)$, $\gkzero(z)$,
$\gkone(z)$, $\gktwo(z)$, $\cJk$ and $\Jk$ are the main results of this
paper and are collected in Appendix~\ref{app:rvFqg}.

\subsection{Technical details}
We have written dedicated {\sc\small Mathematica} parallel codes in order to
apply the whole method to the different contributions.  As already pointed
out in ref.~\cite{Cieri:2019tfv}, the hardest integrals are those to compute
$\tilde{I}_{2}$, which requires the calculation of exact integrals in $z$,
between 0 e $1-f(a)$, where $f(a)$ is defined in eq.~(\ref{eq:def_fa}).

The integrand functions have been classified into five groups, according to
the number of logarithmic and polylogarithmic functions that appear at the
integrand level.  A sample of these integrals is collected in
Appendix~\ref{app:sample_z-int}.
We have integrated $\ord{800}$ integrals in order to compute the expressions
in eqs.~(B.10) and~(B.11) of ref.~\cite{Cieri:2019tfv}, for all the
contributions in eq.~(\ref{eq:I_final_nnlo}).
In general, the integrals require dedicated changes of variables and iterated
integrations by parts, peculiarly for the ones involving polylogarithms and
logarithms to the third power, that turned out to be the most difficult ones.

\subsection{Comments}

Due to the length of the intermediate results, in Appendix~\ref{app:rvFqg} we
report only the final results, i.e.~the functions $\cgkzero(z)$,
$\cgkone(z)$, $\cgktwo(z)$, $\gkzero(z)$, $\gkone(z)$, $\gktwo(z)$, $\cJk$
and $\Jk$ that appear in
eqs.~(\ref{eq:I_final_nnlo})--(\ref{eq:Itilde23_nnlo2}).\footnote{The
  intermediate results are available upon request to the authors.}

In agreement with what is found in ref.~\cite{Cieri:2019tfv}, no odd-power
corrections of $\qtcut/Q = \sqrt{a}$ appear, i.e.~the power expansion of the
real-virtual interference terms for $F$ production in the $qg$ channel is in
$(\qtcut)^2$.

In addition, we can define the
$\left.\hat{G}_{qg}^{\sss(2)}(z)\right|_{\delta(s_2)}$ function, starting
from the integral of the cumulative cross section in
eq.~(\ref{eq:dXS0dqt_V_gq}), as
\begin{equation}
\label{eq:sigma_had_z_one_qg}
\left. \sigma_{qg}^{\sss >(2)}\right|_{\delta(s_2)} = \tau \int_\tau^{1-f(a)} \frac{dz}{z} \,
     {\cal L}_{qg}\!\(\frac{\tau}{z}\) \frac{1}{z}
\left.\sigma_{qg}^{\sss >(2)}(z)\right|_{\delta(s_2)}
   \equiv \tau \int_\tau^1
     \frac{dz}{z} \, {\cal L}_{qg}\!\(\frac{\tau}{z}\)
    \hat{\sigma}^{\sss (0)} \left. \hat{G}_{qg}^{\sss (2)}(z)\right|_{\delta(s_2)} \,,
\end{equation}
and, from the structure of the power corrections we have computed in this
paper, the general form of this function is given by\footnote{The notation
for the expansion of $\left.\hat{G}_{ab}^{(2)}(z)\right|_{\delta(s_2)} $ follows from the number of
powers of $\as$, $\log(a)$ and~$a^\frac{1}{2}$, according to
\begin{equation}
  \label{Ghat_coeff}
 \left. \hat{G}_{ab}^{\sss (2)}(z)\right|_{\delta(s_2)}  =  \sum_{m,r} \,\log^m(a) \,
  \(a^\frac{1}{2}\)^r \hat{G}_{ab}^{\sss(2,m,r)}(z) \,. 
  \nonumber
\end{equation}
}
\begin{eqnarray}
  \label{Ghat_one_expans}
\left.  \hat{G}_{qg}^{\sss(2)}(z)\right|_{\delta(s_2)} &=& {}
  \log^3(a) \, \hat G_{qg}^{\sss(2,3,0)}(z)
  + \log^2(a) \, \hat G_{qg}^{\sss(2,2,0)}(z) +
\log(a)  \, \hat  G_{qg}^{\sss(2,1,0)}(z) + \hat G_{qg}^{\sss(2,0,0)}(z)
\nonumber\\[2mm]
&& {} +   a\log^2(a) \,\hat G_{qg}^{\sss(2,2,2)}(z) +   a\log(a) \,\hat G_{qg}^{\sss(2,1,2)}(z)  +  a \, \hat
G_{qg}^{\sss(2,0,2)}(z)
\nonumber\\[2mm]
&& {}+ \ord{a^\frac{3}{2}\log(a)} ,
\end{eqnarray}
all the other coefficients being zero.

This also agrees with the calculation done in ref.~\cite{Moult:2016fqy},
although the observable is different. In fact, analytic results are therein
obtained for the dominant $\as\,\tau\log(\tau)$ and $\as^2\,\tau\log^3(\tau)$
subleading terms for 0-jettiness~($\tau$) for $q\bar{q}$-initiated
Drell--Yan-like processes.
  
We do not expect this behaviour to be true in general when cuts are applied
to the final-state boson.  This was verified in ref.~\cite{Ebert:2019zkb,
  Ebert:2020dfc}, both for transverse momentum and $N$-jettiness. In fact,
power corrections proportional to $\sqrt{a}$ and $\sqrt{\tau}$ are found
therein.

\section{Conclusions}
\label{sec:conclusions}

In this paper we considered the production of a vector boson $F$ ($Z$,
$W^\pm$, $\gamma^*$) at next-to-next-to-leading order in the strong coupling
constant $\as$.  We imposed a transverse-momentum cutoff, $\qtcut$, on the
vector boson and we computed, up to the second power of $\qtcut$, the power
corrections for the $qg$-initiated real-virtual contributions to the
cumulative cross section, and for other contributions from double-real
radiation, leftover of the subtraction scheme, having  the same kinematics,
i.e.~$F$ + 1 parton.

Although we studied Drell--Yan-type $F$ boson production, the procedure we
followed is general and can be applied to other similar cases, up to any
order in the powers of $\qtcut$, as illustrated in our previous
paper~\cite{Cieri:2019tfv}.

We presented analytic results for the power corrections in $\qtcut$
and we found that the logarithmic terms in $\qtcut$ show up at most to the third
power in the power-correction contributions, as expected, and that 
no odd-power corrections in $\qtcut$ appear. This is in agreement with known
results in the literature at a lower order in $\as$, i.e.~next-to-leading,
and with what we found in ref.~\cite{Cieri:2019tfv} where we computed the
power corrections at next-to-leading order up to $(\qtcut)^4$.  We do not
expect this to be true in general when cuts are applied to the final state.

The knowledge of the power terms is crucial for understanding both the
non-trivial behaviour of cross sections in the infrared limit, and the
resummation structure at subleading orders.  In addition, within the
$\qt$-subtraction method, the knowledge of the power terms helps in reducing
the cutoff dependence of the cross sections.

The result presented in this paper is the first step towards the full
calculation of the power corrections of vector-boson production at NNLO. Work
is ongoing to compute these corrections for the $qq$-initiated real-virtual
contributions and for the double-real radiation contributions too.

\section*{Acknowledgments}

We thank L.~Cieri for useful discussions at the early stages of the paper.

\appendix

\section{Samples of integrals}
\label{app:sample_z-int}
According to the procedure first presented in ref.~\cite{Cieri:2019tfv}, in
order to compute the power corrections, one has to perform two integrations of
the differential cross sections: an integration in $\qt$, in general easy to
perform, and an integration in $z$ from 0 to $1-f(a)$.
This second integration turned out to be challenging for some integrand
functions.

We have classified the integrand functions into five groups, according to the
number of logarithms and polylogarithms appearing in the expressions.  We
present here a sample of integrands for each group.

\subsection{Integrand classification}
Defining
\begin{equation}
   r(z) \equiv \sqrt{(1-z)^2-4az} \,,
\end{equation}
we have groups containing:
\begin{enumerate}
\item one logarithm: 
  \begin{eqnarray}
    && \int_0^{1-f(a)} d z \, z^n \,
    \log \lq (1-z)\,\frac{1+z \pm r(z)}{1-z \pm r(z)} \rq 
    \\
    && \int_0^{1-f(a)} d z \, z^n \, \frac{r(z)}{1+z \pm r(z)} \,
    \log \lq \frac{2z}{1-z \pm r(z)}\rq
  \end{eqnarray}

\item two logarithms: 
   \begin{eqnarray}
&& \int_0^{1 -f(a)}  dz \,
     z^n  \, r(z) \log(z) \log \left[\frac{1-z -r(z)}{2(1-z)}\right]   
\\ 
    && \int_0^{1-f(a)} d z \, z^n  \, r(z) \, \log^2 \lq \frac{2z}{1-z \pm r(z)} \rq
   \end{eqnarray}

 \item three logarithms: 
   \begin{eqnarray}
&& \int_0^{1 -f(a)}  dz \,
z^n\, \log ^2\left[\frac{1-z -r(z)}{1-z+r(z)}\right]
\log \left[(1-z) \frac{1+z + r(z)}{1-z+r(z)}\right]
\\
&&  \int_0^{1 -f(a)}  dz \,
 z^n \, \log(z)\log^2\lq \frac{1-z \pm r(z)}{2(1-z)} \rq
\\ 
    && \int_0^{1-f(a)} d z \, z^n \, \log^3 \lq \frac{1\mp z \pm r(z)}{2z}\rq    
    \\
    && \int_0^{1-f(a)} d z \, z^n\, \log(z) \,
    \log\frac{1-z\pm r(z)}{2\,(1-z)} \,
    \log\frac{1-z \mp r(z)}{2z}
 \phantom{aaaaaa}
   \end{eqnarray}

  \item one polylogarithm of order 2:
    \begin{eqnarray}
      &&  \int_0^{1-f(a)} d z \,  z^n \, r(z) \,
      \litw\lq \frac{2z}{1+z\pm r(z)} \rq
      \\
      && \int_0^{1-f(a)} d z \, z^n \, \log\lq \frac{1-z \pm r(z)}{1-z \mp r(z)}\rq \,
      \litw \lq -z\,\frac{1-z \pm r(z)}{1-z \mp r(z)} \rq
    \end{eqnarray}

  \item  one polylogarithm of order 3:
    \begin{eqnarray}
      && \int_0^{1-f(a)} d z \, z^n \, \lith\lq \frac{2z}{1 +  z  \pm r(z)}
      \rq 
      \\
      && \int_0^{1-f(a)} d z \, z^n \,\lith\lq -z \,\frac{1-z\pm r(z)}{1-z
        \mp r(z)}
      \rq   
    \end{eqnarray}

\end{enumerate}
where $ n=1,\,\ldots,\, 4$.

\subsection{Sample of integral expansion}
After the $z$ integration, the results are functions of $a$ only, and
have to be expanded around $a=0$.
A sample of these expansions is given in the following:
\begin{itemize}
\item[-] {\bf Example 1}
\begin{eqnarray}
&& \int_0^{1 -f(a)}  dz \,
z \log ^2\left[\frac{1-z -r(z)}{1-z+r(z)}\right]
\log \left[(1-z) \frac{1+z + r(z)}{1-z+r(z)}\right]
\nonumber\\[2mm]
&& \hspace{1cm} = \frac{1}{2} \,a \log^3(a)
  +2\, a \log^2(a) +
  \left(\frac{17}{2} + \pi ^2\right) a \log(a)
  +\lq 48 -32\, C-16 \log 2\rq\sqrt{a}
 \nonumber\\[2mm]
&& \hspace{1.4cm} 
+ \lq 9 \,\zeta (3)-\frac{15}{4}+\frac{4 }{3}\pi^2 +8 \log2 \rq a
 + \ord{a^{\frac{3}{2}}} ,
\end{eqnarray}
where $C$ is the Catalan constant defined by
\begin{equation}
 C = \sum_{n=0}^\infty \frac{(-1)^n}{(2n+1)^2} = \frac{1}{1^2} -
 \frac{1}{3^2} + \frac{1}{5^2}  - \frac{1}{7^2} + \ldots \approx
 0.915965594\ldots 
\end{equation}

\item[-] {\bf Example 2}
\begin{eqnarray}
&& \int_0^{1 -f(a)}  dz \,
z^3 \log ^2\left[\frac{1-z -r(z)}{1-z+r(z)}\right]
\log \left[(1-z) \frac{1+z + r(z)}{1-z+r(z)}\right]
\nonumber\\[2mm]
&& \hspace{1cm} = \frac{5}{6} a \log ^3(a) +\frac{71}{12} \,a \log ^2(a)
+\left(\frac{1721}{72}+\frac{5}{3} \pi^2\right) a \log (a)
\nonumber\\[2mm]
&& \hspace{1.4cm}
+ \lq 48 -32\, C -16 \log2\rq \sqrt{a}
\nonumber\\[2mm]
&& \hspace{1.4cm}
+ \left[ 16 \,\zeta(3)+\frac{8711}{864}+\frac{71 }{18}\pi^2
+ 16 \log2\right] a + \ord{a^{\frac{3}{2}}} ,
\end{eqnarray}

\item[-] {\bf Example 3}
\begin{eqnarray}
&& \int_0^{1 -f(a)}  dz \,
z  \, r(z) \log(z) \log \left[\frac{1-z -r(z)}{2(1-z)}\right]
\nonumber\\[2mm]
&& \hspace{1cm} =
-\frac{5}{36}\log (a) +\frac{\pi^2}{18}-\frac{55}{108}
+\left(\frac{\pi^2}{3}-\frac{5}{2}\right) a \log (a)
+ \left(\frac{5 }{6}\pi^2 -\frac{25}{4}\right) a
+ \ord{a^{\frac{3}{2}}} ,
\nonumber
\\
\end{eqnarray}

\item[-] {\bf Example 4}
\begin{eqnarray}
&&  \int_0^{1 -f(a)}  dz \,
z   \log(z) \log \left[\frac{1-z + r(z)}{2(1-z)}\right]
= 
a \left[1-\frac{\pi ^2}{12}+2 \log^22-2 \log 2\right]
+ \ord{a^{\frac{3}{2}}} ,
\nonumber
\\
\end{eqnarray}


\item[-] {\bf Example 5}
\begin{eqnarray}
&&  \int_0^{1 -f(a)}  dz \,
 z  \log(z)\log^2\lq \frac{1-z-r(z)}{2(1-z)} \rq
\nonumber\\[2mm]
&& \hspace{1cm} =
\frac{1}{2} a \log ^2(a)-\frac{\log ^2(a)}{4}+a \left(2+\frac{3 }{4}\pi^2
+2 \log^22-2 \log 2\right)
\nonumber\\[2mm]
&& \hspace{1.4cm}
+\left(\frac{2}{3} \pi^2 -4\right) a
\log (a)+\left(\frac{\pi^2}{3}-\frac{7}{2}\right) \log (a)
+\frac{2}{3} \pi^2 -\frac{59}{8}
+ \ord{a^{\frac{3}{2}}} .
\nonumber\\
\end{eqnarray}

\end{itemize}
We note that the intermediate integrals contain $\log(2)$ and $\sqrt{a}$
terms, and also terms proportional to the Catalan constant $C$.  Despite
this, once recombined to compose the whole behaviour of the physical cross
section, all these terms disappear from the final answer, as illustrated in
Appendix~\ref{app:rvFqg}.  Something similar happened for the results at NLO
we presented in ref.~\cite{Cieri:2019tfv}.

\section{Final results}
\label{app:rvFqg}
In this appendix we collect the results for the NNLO power corrections, up to
order $a$ in the transverse-momentum cutoff. The results refer to the
$\delta(s_2)$ contribution of the $qg$-initiated channel to the inclusive
cross section for the production of a vector boson $F$, i.e.  the
$\cgkzero(z)$, $\cgkone(z)$, $\cgktwo(z)$, $\gkzero(z)$, $\gkone(z)$,
$\gktwo(z)$, $\cJk$ and $\Jk$ functions in
eqs.~(\ref{eq:Itilde1_nnlo1})--(\ref{eq:Itilde23_nnlo2}).

In the following, we need $l(z)$, defined in eq.~(\ref{eq:def_lz}), and its
first derivative
\begin{equation}
 l^{(1)}(z) \equiv \frac{d }{dz}l(z) = -\frac{1}{z^2}\,\Lum\(\frac{\tau}{z}\)
  -\frac{\tau}{z^3}\,\Lum^{(1)}\(\frac{\tau}{z}\),
\end{equation}
both evaluated in $z=1$.
For sake of brevity, we introduce the following notation
\begin{eqnarray}
  \lumn &\equiv& l(1) = \Lum\(\tau\) ,
  \\
  \lumnn &\equiv& l^{(1)}(1) = -\Lum\(\tau\)-\tau\Lum^{(1)}\(\tau\)  .
\end{eqnarray}
The renormalisation and factorisation scales are indicated with $\mur$ and
$\muf$, respectively, and $\APreg_{qg}(z) $ is the zeroth-order
Altarelli--Parisi splitting function, defined as
\begin{equation}
P_{qg}(z) =  \TR \lq  2 z^2 -2 z+ 1 \rq \equiv \TR \, \APreg_{qg}(z) \,.
\end{equation}
In addition, we recall the definition of $a$ in eq.~(\ref{eq:a_def}):
$    a=\(\qtcut\)^2/Q^2$.

\subsection[$A^{qg}$]{$\boldsymbol{A^{qg}}$}
\begin{eqnarray} 
  g_{\sss 0}^{\sss A}(z) &=&  
   \APreg_{qg}(z)\lq  -\log(a) + \log\frac{(1-z)^2}{z} \rq
 + \frac{1}{2}\,(1+3z)(1-z)  + \ord{a^\frac{3}{2}\log(a)}
\end{eqnarray}
\begin{eqnarray} 
  g_{\sss 1}^{\sss A}(z) &=&  - z\,(1+3z)\, a  +   \ord{a^{\frac{3}{2}}\log(a)} 
\end{eqnarray}
\begin{eqnarray} 
  g_{\sss 2}^{\sss A}(z) &=&  
   - 2z\,\APreg_{qg}(z)\, a  + \ord{a^{\frac{3}{2}}\log(a)} 
\end{eqnarray}
%
\begin{eqnarray}
  {\cal J}^{\sss A}_{\sss 23} &=& 
   -\(\lumn+\lumnn\)  a \log(a)  -
  \(\frac{3}{2}\lumn  +  \frac{10}{3}\lumnn \)   a
  + \ord{a^{\frac{3}{2}}\log(a)} 
\end{eqnarray}

\subsection[$B_1^{qg}$: $\CA$ coefficient]
           {$\boldsymbol{B_1^{qg}}$: $\boldsymbol{\CA}$ coefficient}

\begin{eqnarray} 
   ^{\sss\CA\!}g_{\sss 0}^{\sss B_1}(z) &=& \frac{1}{6}\,\APreg_{qg}(z)\log^3(a)
  -\APreg_{qg}(z)\log(z)\log^2(a)
  \nonumber\\[2mm]
  &&{} + \lq   z - \APreg_{qg}(z)  \( \log^2\frac{z}{1-z} -
  2 \log(1-z)\log (z)  + \frac{7}{6}\pi^2 \) \rq \log (a)
  \nonumber\\[2mm]
  &&{} + \APreg_{qg}(z) \( \frac{2}{3} \log^3(1-z)
 - \frac{1}{6}\log^3(z) + \log(1-z)\log^2(z) \)
  \nonumber\\[2mm]
  &&{} - \( 6z^2 - 5z + \frac{5}{2} \) \log^2(1-z)\log(z)
    - \( \frac{15}{2}\,z^2 - 5z + 1 \) \log(1-z)\log(z)
\nonumber\\[2mm]
  &&{} 
  + \( \frac{21}{4}\,z^2 - \frac{3}{2}\,z - \frac{1}{4} \) \log^2(z)
  + \(3z^2 - 4z + 1 \) \log^2(1-z) 
    \nonumber\\[2mm]
  &&{} + \( \frac{14}{3}\,\pi^2\,z^2 + \frac{9}{2}\,z^2 -
  \frac{9}{2}\,\pi^2\,z - 8z + \frac{9}{4}\,\pi^2
  + \frac{3}{2}  \) \log(1-z)
  \nonumber\\
  &&{} -\left(\frac{4}{3}\,\pi^2\,z^2 + \frac{3}{2}\,z^2 -
  \frac{3}{2}\,\pi^2\,z - 3z + \frac{3}{4}\,\pi^2 - \frac{1}{2} 
  \) \log(z)
  \nonumber\\[2mm]
  &&{} - \( \frac{3}{2}\,z^2 + 3z - 1 + \frac{1}{2} (2z-1)\,\log\frac{z}{1-z}
  \)
  \Big(\Litw(1-z)-\Litw(z)\Big)
  \nonumber\\[2mm]
  &&{} - 2(2z-1) \, \Lith(1-z) - \(4z^2 - 2z + 1\) \Lith(z)
  \nonumber\\[2mm]
  &&{} +  \zeta(3) \( 4z^2 - 2z + 1\)
  - \frac{15}{4}\,\pi^2\,z^2 -
  \frac{13}{4}\,z^2  + \frac{23}{6}\,\pi^2\,z
  + \frac{9}{2}\,z - \frac{2}{3}\,\pi^2 -
  \frac{5}{4}
  \nonumber\\[2mm]
  &&{} 
 + \lq \frac{11}{6}\, \APreg_{qg}(z) \( -\log (a)
 + \log\frac{(1-z)^2}{z} \)
 + \frac{11}{12}\, (1-z)\, (3z+1) \rq \log\frac{\mur^2}{Q^2}
  \nonumber\\[2mm]
  &&{}  - 2\,(1+z)  \, a\log^2(a)
  \nonumber\\[2mm]
  &&{} + \lq \frac{39}{8}\,z - \frac{45}{16} 
   + (11z+10)\,\log(1-z) - (16z+19)\,\log(z) \rq a\log(a)
  \nonumber\\[2mm]
  &&{} + \lq - \( \frac{13}{8}\,z^3 + 9z^2 + \frac{27}{16}\,z
  - \frac{1}{16} \) \log (1-z) + \( \frac{5}{2}\,z + 2 \) \log ^2(1-z)
  \right.\nonumber\\[2mm]
  &&{} \hspace{7.5mm} + \( \frac{13}{8}\,z^3 + 9\,z^2 -\frac{63}{8}\,z -
  \frac{11}{2} \) \log(z) - \( \frac{19}{2}\,z + 5 \) \log^2(z)
  \nonumber\\[2mm]
  &&{} \hspace{7.5mm} - \( z + 9 \) \log(1-z)\log(z)
  + 2\(1+z\)  \Big( \Litw(1-z)-\Litw(z)\Big)
  \nonumber\\[2mm]
  &&{} \hspace{7.5mm} \left. +\,  \frac{23}{8} z^2 - \frac{165}{16} z +
  \frac{\pi^2}{3} z - 3 + \frac{5}{6} \pi^2  \rq a
  + \ord{a^{\frac{3}{2}}\log(a)}
\end{eqnarray}

\begin{eqnarray} 
  ^{\sss\CA\!}g_{\sss 1}^{\sss B_1}(z) &=&
   \lq -2z^3 - \frac{1}{2}\,z^2 + z + \frac{1}{2} \rq a\log^2(a)
  \nonumber\\[2mm]
  &&{} + \biggl[ 
  \(8z^3+12z^2-8z-4\)\log(1-z) + \(-4z^3-11z^2+6z+1\)\log(z) 
  \nonumber\\[2mm]
  &&{} \hspace{7.5mm} \left.
  +\, 4z^3 - \frac{43}{8}\,z^2 - \frac{75}{16}\,z + \frac{61}{16} \rq a\log(a)
  \nonumber\\[2mm]
  &&{} + \biggl[ \(-8z^3+\frac{1}{2}\,z^2+\frac{15}{2}\,z-8\)\log^2(1-z) +
  \(6z^3-10z^2+\frac{11}{2}\,z\)\log^2(z)
  \nonumber\\[2mm]
  &&{} \hspace{7.5mm} + \(12z^3-\frac{23}{2}\,z^2-2z+15\)\log(1-z)\log(z)
  \nonumber\\[2mm]
  &&{} \hspace{7.5mm} - \(1+z\) \( \frac{13}{8}\,z^3 + \frac{31}{4}\,z^2 - \frac{177}{16}\,z
  + \frac{65}{16} \) \log(1-z)
  \nonumber\\[2mm]
  &&{} \hspace{7.5mm} + \(\frac{9}{2}\,z^2-z-2\) \Big(\Litw(1-z)-\Litw(z)\Big)
  \nonumber\\[2mm]
  &&{} \hspace{7.5mm} 
  + \( \frac{13}{8}\,z^4 + \frac{43}{8}\,z^3 - \frac{29}{8}\,z^2 -
  \frac{43}{8}\,z + \frac{17}{2} \) \log(z)
  \nonumber\\[2mm]
  &&{} \hspace{7.5mm} 
  +  2\pi^2\,z^3 - \frac{63}{16}\,z^3 - \frac{79}{12}\,\pi^2\,z^2
  - \frac{11}{4}\,z^2 - \frac{\pi^2}{2}\,z + \frac{79}{16}\,z +
  \frac{4}{3}\,\pi^2 + 4 
  \nonumber\\[2mm]
  && \hspace{7.5mm} \left.
 {} - \frac{11}{6} z \(3z+1\) \log\frac{\mur^2}{Q^2}  \rq a + \ord{a^{\frac{3}{2}}\log(a)}
\end{eqnarray}

\begin{eqnarray} 
 ^{\sss\CA\!}g_{\sss 2}^{\sss B_1}(z) &=&
\lq \frac{1}{2} (1-z) (z+1) \left(4 z^2-6 z+3\right) \rq a \log ^2(a)
\nonumber\\[2mm]
&&{} +  \bigg[  (z-1) \left(4 z^3- \frac{21}{4} z^2-2 z+3\right)
  +2 \left(4 z^4-4z^3-3 z^2+7 z-3\right)\log (1-z)
\nonumber\\[2mm]
&&{} \hspace{7.5mm} 
+\left(-4 z^4+10 z^3+2 z^2-29 z+18\right) \log (z)  \bigg] a \log (a)
\nonumber\\[2mm]
&&{} + \bigg[ \left(12 z^4- 25 z^3+\frac{43}{2} z^2+2 z-6\right) \log (1-z) \log (z)
\nonumber\\[2mm]
&&{} \hspace{7.5mm}
-2 \left(4 z^4-4 z^3-3 z^2+7 z-3\right) \log ^2(1-z)
\nonumber\\[2mm]
&&{} \hspace{7.5mm}
+ \left(6 z^4-5 z^3- \frac{3}{2} z^2-5 z+5\right) \log ^2(z)
\nonumber\\[2mm]
&&{} \hspace{7.5mm}
+\frac{1}{4} \left(16 z^4-11 z^3-32 z^2+31 z-12\right) \log (z)
\nonumber\\[2mm]
&&{} \hspace{7.5mm}
+(1-z) \left(8 z^3-4 z^2-\frac{25}{4} z+6\right)\log (1-z)
\nonumber\\[2mm]
&&{} \hspace{7.5mm}
+\frac{z^2}{2} \!\(2z-1\) \! \Big(\Litw(1-z)-\Litw(z)\Big)
+\frac{\pi^2}{12}\! \left(24 z^4-94 z^3+45 z^2+24 z-26\right)
\nonumber\\[2mm]
&&{} \hspace{7.5mm}
+ \frac{1}{16} \left(-109 z^4+132 z^3+83 z^2-122 z+48\right)
-\frac{11}{3} z \, \APreg_{qg}(z) \log \frac{\mur^2}{Q^2} \bigg] a
\nonumber\\[2mm]
&&{} 
+ \ord{a^{\frac{3}{2}}\log(a)}
\end{eqnarray}

\begin{eqnarray}
  ^{\sss\CA\!}{\cal J}^{\sss B_1}_{\sss 23} &=&
 \frac{1}{6}  \lq \lumn + \lumnn \rq a\log^3(a)
  + \frac{1}{4} \lq 15 \,\lumn - \frac{11}{3}\,\lumnn \rq a\log^2(a) 
  \nonumber\\[2mm]
  &&{} + \lq \( -\frac{11}{6}\,\log\frac{\mur^2}{Q^2} - \frac{5}{3}\,\pi^2 -
  \frac{43}{8} \) \lumn \right.
  \nonumber\\[2mm]
  &&{} \hspace{7.5mm} \left.
  + \(-\frac{11}{6}\,\log\frac{\mur^2}{Q^2} - \frac{5}{3}\,\pi^2 
  + \frac{79}{6} \) \lumnn \rq a\log(a) 
  \nonumber\\[2mm]
  &&{} + \lq \( -\frac{11}{4}\,\log\frac{\mur^2}{Q^2} - 5\zeta(3) - \frac{11}{3}\,\pi^2 +
  \frac{897}{32} \) \lumn \right.
  \nonumber\\[2mm]
  &&{} \hspace{7.5mm} \left.
  + \(-\frac{55}{9}\,\log\frac{\mur^2}{Q^2} - 5\zeta(3) - \frac{185}{36}\,\pi^2 
  + \frac{565}{192} \) \lumnn \rq a + \ord{a^{\frac{3}{2}}\log(a)}\phantom{aaaaaaaa}
\end{eqnarray}

\subsection[$B_1^{qg}$: $\CF$ coefficient]
           {$\boldsymbol{B_1^{qg}}$: $\boldsymbol{\CF}$ coefficient}

\begin{eqnarray} 
   ^{\sss\CF\!}g_{\sss 0}^{\sss B_1}(z) &=&
  \APreg_{qg}(z)\log\frac{z}{1-z}\log^2(a)
  \nonumber\\[2mm]
  &&{} + \bigg[  16z^2   - 17z  + 8  + \APreg_{qg}(z) \bigg(
    \frac{11}{6}\pi^2+ \Litw(z) -  \Litw(1-z)  
  \nonumber\\[2mm]
  &&{} 
  \hspace{7.5mm}+ 2  \log^2\frac{z}{1-z} + \log^2(1-z)  -\log(1-z)\log (z) \bigg) \bigg] \log (a)
  \nonumber\\[2mm]
  &&{} -2\APreg_{qg}(z) \( \log^3(1-z)
  - \frac{2}{3}\log^3(z)  + 2\log(1-z)\log^2(z)\)
  \nonumber\\[2mm]
  &&{} + \( 8z^2 - 6z + 3 \) \log^2(1-z)\log(z) 
  \nonumber\\[2mm]
  &&{}   + \( -\frac{9}{2}\,z^2 + 7z - \frac{5}{2} \) \log^2(1-z) 
  + \( -6z^2 + 3z - 2 \) \log^2(z) 
  \nonumber\\[2mm]
  &&{} + \( -7\pi^2\,z^2 - 42z^2 + \frac{20}{3}\,\pi^2\,z
  + 48z - \frac{10}{3}\,\pi^2 - 20  \) \log(1-z)
  \nonumber\\[2mm]
  &&{} + \( \frac{8}{3}\,\pi^2\,z^2 + \frac{47}{2}\,z^2 -
  3\pi^2\,z - 24z + \frac{3}{2}\,\pi^2 + 9 \) \log(z)
  \nonumber\\[2mm]
  &&{} + \( \frac{21}{2}\,z^2 - 10z + \frac{9}{2} \) \log(1-z)\log(z)
  + \lq \frac{3}{2}\,z^2 + 4z - \frac{1}{2} \right.
  \nonumber\\[2mm]
  &&{} \hspace{7.5mm}      + \(6z^2 - 8z + 4\)\log(1-z)
  - \(4z^2 - 6z + 3\) \log(z) \biggr] \Big(\Litw(1-z)-\Litw(z)\Big)
  \nonumber\\[2mm]
  &&{} - 8(1-z)^2 \, \Lith(1-z) + 2\(2z - 1\) \Lith(z)
  +  \frac{27}{4}\,\pi^2\,z^2 + \frac{37}{2}\,z^2
  \nonumber\\[2mm]
  &&{}  - 4z\,\zeta(3) - \frac{23}{3}\,\pi^2\,z
  - \frac{33}{2}\,z + 2\zeta(3) + \frac{7}{4}\,\pi ^2 - 2
  + \frac{1}{2}\,z \, a\log^2(a)
  \nonumber\\[2mm]
  &&{} + \lq -\frac{73}{8}\,z + \frac{25}{4} 
   - (21z+20)\,\log(1-z) + (33z+38)\,\log(z) \rq a\log(a)
  \nonumber\\[2mm]
  &&{} + \lq  \( \frac{19}{8}\,z^3 + 14z^2 + 3z
  - 2 \) \log (1-z) - 12 \( 1+z \) \log^2(1-z)
  \right.\nonumber\\[2mm]
  &&{} \hspace{7.5mm} + \( -\frac{19}{8}\,z^3 - 14\,z^2 + \frac{141}{8}\,z +
  \frac{31}{2} \) \log(z) + \( 16z + 8 \) \log^2(z)
  \nonumber\\[2mm]
  &&{} \hspace{7.5mm} + \( -\frac{9}{2}\,z + 14 \) \log(1-z)\log(z)
  - \( \frac{7}{2}\,z + 4 \) \Big(\Litw(1-z)-\Litw(z)\Big)
  \nonumber\\[2mm]
  &&{} \hspace{7.5mm} \left. -\,  \frac{43}{8}\,z^2 + \frac{175}{8}\,z +
  \frac{23}{12}\,\pi^2\,z + \frac{47}{8} + \pi^2  \rq a + \ord{a^{\frac{3}{2}}\log(a)}
\end{eqnarray}

\begin{eqnarray} 
   ^{\sss\CF\!}g_{\sss 1}^{\sss B_1}(z) &=&
  \lq 4z^3 - \frac{1}{2}\,z^2 - \frac{5}{2}\,z + 3 \rq a\log^2(a)
  \nonumber\\[2mm]
  &&{} + \biggl[ 
  \(-16z^3-27z^2+15z+8\)\log(1-z) + \(8z^3+19z^2-13z-2\)\log(z) 
  \nonumber\\[2mm]
  &&{} \hspace{7.5mm} \left.
  -\, 8z^3 + \frac{13}{8}\,z^2 + \frac{69}{8}\,z - \frac{25}{4} \rq a\log(a)
  \nonumber\\[2mm]
  &&{} + \biggl[ \(16z^3-5z^2-15z+24\)\log^2(1-z) +
  \(-8z^3+19z^2-8z\)\log^2(z)
  \nonumber\\[2mm]
  &&{} \hspace{7.5mm} + \(-22z^3+\frac{37}{2}\,z^2+\frac{11}{2}\,z-26\)\log(1-z)\log(z)
  \nonumber\\[2mm]
  &&{} \hspace{7.5mm} + \( \frac{19}{8}\,z^4 + \frac{125}{8}\,z^3 + 2z^2 -
  12z + 8 \) \log(1-z)
  \nonumber\\[2mm]
  &&{} \hspace{7.5mm} 
  + \( -\frac{19}{8}\,z^4 - \frac{61}{8}\,z^3 + \frac{51}{8}\,z^2 +
  \frac{89}{8}\,z - \frac{37}{2} \) \log(z)
  \nonumber\\[2mm]
  &&{} \hspace{7.5mm} + \(2z^3-\frac{25}{2}\,z^2+\frac{5}{2}\,z+4\)
  \Big(\Litw(1-z)-\Litw(z)\Big)
  \nonumber\\[2mm]
  &&{} \hspace{7.5mm}
  - \frac{13}{3}\,\pi^2\,z^3 + \frac{125}{16}\,z^3 + \frac{169}{12}\,\pi^2\,z^2
  + \frac{599}{16}\,z^2 + \frac{5}{4}\,\pi^2\,z - \frac{51}{16}\,z
   \nonumber\\[2mm]
 &&{} \hspace{7.5mm} \left.
  -  \frac{17}{3}\,\pi^2 - \frac{55}{8} \rq a 
+ \ord{a^{\frac{3}{2}}\log(a)}
\end{eqnarray}

\begin{eqnarray} 
  ^{\sss\CF\!}g_{\sss 2}^{\sss B_1}(z) &=& \( 4 z^4-4 z^3-3 z^2+7 z-3 \) a \log^2(a)
  \nonumber\\[2mm]
  &&{}  
+ \bigg[ 4 \left(2 z^4-6 z^3+14 z-9\right) \log (z) + \frac{1}{4} (1-z)
\left(32 z^3-35 z^2-13 z+24\right)
 \nonumber\\[2mm]
  &&{} \hspace{7.5mm} 
-2 \left(8 z^4-6 z^3-8 z^2+15 z-6\right) \log (1-z) \bigg] a \log(a)
  \nonumber\\[2mm]
  &&{}
+ \bigg[ 2 z \left(z^3-4 z^2+3 z-1\right) \,\Big(\Litw(1-z)-\Litw(z)\Big)
\nonumber\\[2mm]
  &&{} \hspace{7.5mm} 
 +2 \left(8 z^4-6
 z^3-8 z^2+15 z-6\right) \log ^2(1-z)
\nonumber\\[2mm]
  &&{} \hspace{7.5mm}
 +\left(-8 z^4+10 z^3-z^2+10 z-8\right)
 \log ^2(z)
\nonumber\\[2mm]
  &&{} \hspace{7.5mm}
 -2 \left(11 z^4-24 z^3+21 z^2+2 z-6\right) \log (z) \log
 (1-z)
 \nonumber\\[2mm]
  &&{} \hspace{7.5mm} 
 +4 (z-1) (z+1) \left(4 z^2-6 z+3\right) \log(1-z)
\nonumber\\[2mm]
  &&{} \hspace{7.5mm}
+\frac{1}{4} \left(-32 z^4-z^3+74 z^2-65 z+12\right) \log (z)
\nonumber\\[2mm]
&&{} \hspace{7.5mm}
+\frac{\pi^2}{3} \left(-13 z^4+42 z^3-15 z^2-17 z+14\right)
\nonumber\\[2mm]
  &&{} \hspace{7.5mm} 
  + \frac{1}{16} \left(211 z^4+208 z^3-566 z^2+467 z-96\right) \bigg] a
+ \ord{a^{\frac{3}{2}}\log(a)} \phantom{aaaaa}
\end{eqnarray}

\begin{eqnarray}
  ^{\sss\CF\!}{\cal J}^{\sss B_1}_{\sss 23} &=&
  \lq -\frac{3}{4}\,\lumn + \frac{7}{2}\,\lumnn \rq a\log^2(a) 
  \nonumber\\[2mm]
  &&{} + \lq \( \frac{8}{3}\,\pi^2 + \frac{429}{16} \) \lumn 
  + \( \frac{8}{3}\,\pi^2 - \frac{821}{72} \) \lumnn \rq a\log(a) 
  \nonumber\\[2mm]
  &&{} + \lq \( 8\zeta(3) + \frac{\pi^2}{3} +
  \frac{4627}{192} \) \lumn 
  + \( 8\zeta(3) + \frac{64}{9}\,\pi^2 
  + \frac{75667}{1728} \) \lumnn \rq a
  + \ord{a^{\frac{3}{2}}\log(a)}
 \nonumber\\  
\end{eqnarray}

\subsection[$B_2^{qg}$]{$\boldsymbol{B_2^{qg}}$}
\begin{eqnarray} 
  g_{\sss i}^{\sss B_2}(z) &=&  -\frac{1}{3}\log\frac{\mur^2}{Q^2}\;g_{\sss i}^{\sss
    A}(z) \hspace{1.5cm} i=0,1,2
  \\[2mm]
 {\cal J}^{\sss B_2}_{\sss 23} &=& -\frac{1}{3} \log\frac{\mur^2}{Q^2}\; {\cal J}^{\sss A}_{\sss 23}
\end{eqnarray}

\subsection[$B_3^{qg}$]{$\boldsymbol{B_3^{qg}}$}
\begin{eqnarray} 
  g_{\sss 0}^{\sss B_3}(z) &=&  
   \frac{1}{2} \,z \,\log^2(z)  +2z \( 1-z \) \log\frac{z}{1-z}  
  \nonumber\\[2mm]
  &&{} - \frac{1}{2}\,z \,\log(1-z)\log(z)
  - \frac{1}{2}\,z\, \Big(\Litw(1-z)-\Litw(z)\Big)
  \nonumber\\[2mm]
  &&{} -  \frac{1}{12}\,\pi^2\,z - z + 1 
   + \biggl[  \( 1+z \) \log (1-z)  \biggr] a + \ord{a^\frac{3}{2}\log(a)}
\end{eqnarray}
\begin{eqnarray} 
  g_{\sss 1}^{\sss B_3}(z) &=&  z \, a\log(a)
  + \biggl[  (z-1)\(2z^2+2z+1\) \,\log(1-z) 
    - 2z^3\,\log(z) - z \biggr] a
 \nonumber\\[2mm]
  &&{}
  + \ord{a^{\frac{3}{2}}\log(a)}
\end{eqnarray}

\begin{eqnarray} 
  g_{\sss 2}^{\sss B_3}(z) &=&  \(1-z\) z^2\big[ 2  +  \log (z) \big] a 
  + \ord{a^{\frac{3}{2}}\log(a)}
\end{eqnarray}

%
\begin{eqnarray}
  {\cal J}^{\sss B_3}_{\sss 23} &=&
  -\frac{1}{2}\,\lumn \, a\log^2(a)
  -\lumn \, a\log(a) 
  + \lq \(  \frac{\pi^2}{6} +  \frac{1}{4} \) \lumn 
  - \frac{5}{9} \, \lumnn \rq a + \ord{a^{\frac{3}{2}}\log(a)}
  \nonumber\\
\end{eqnarray}

\subsection[$C_1^{qg}$: $\CA$ coefficient]
           {$\boldsymbol{C_1^{qg}}$: $\boldsymbol{\CA}$ coefficient}

\begin{eqnarray} 
   ^{\sss\CA\!}g_{\sss 0}^{\sss C_1}(z) &=& -\frac{1}{6}\,\APreg_{qg}(z)\log^3(a)
  +\APreg_{qg}(z)\log(1-z)\log^2(a)
  \nonumber\\[2mm]
  &&{} - \APreg_{qg}(z)  \lq \frac{\pi^2}{6} + 2\log^2(1-z)  \rq \log (a)
  \nonumber\\[2mm]
  &&{} + \frac{1}{3}\,\APreg_{qg}(z)   \( 4\log^3(1-z)
  +  \pi^2  \log\frac{1-z}{z} - \frac{1}{2}\,\log^3(z)\)
  \nonumber\\[2mm]
  &&{} 
  + \( -\frac{3}{4}\,z^2 + \frac{1}{2}\,z + \frac{1}{4} \) \log^2(z) 
   + \( - \frac{1}{2}\,z^2 + z  - \frac{1}{2}  \) \log(z)
  \nonumber\\[2mm]
  &&{} 
  + \frac{\pi^2}{6}\( 1 + 2z -3z^2 \)
  - \lq \frac{3}{2}\,z + \frac{3}{2} + z \log(z) \rq a\log(a)
  \nonumber\\[2mm]
  &&{} + \lq 2z \log(1-z)\log(z) -z \log^2(z)- \( z + \frac{3}{2} \) \log(z)
   \rq a  + \ord{a^{\frac{3}{2}}\log(a)}
  \nonumber\\
\end{eqnarray}

\begin{eqnarray} 
   ^{\sss\CA\!}g_{\sss 1}^{\sss C_1}(z) &=&
 -\frac{z}{2}  \( 3 z  + 1 \) a\log^2(a)
  \nonumber\\[2mm]
  &&{} + \lq  z \(3z+1\) \(  \log \frac{(1-z)^2}{z}
  + \frac{1}{2} \) +\frac{3}{2} \rq a\log(a)
  \nonumber\\[2mm]
  &&{} + \biggl[2 z \(3z+1\) \( \log(1-z)\log(z) - \log^2(1-z) - \frac{1}{2}\log^2(z)\)
  \nonumber\\[2mm]
  &&{} \hspace{7.5mm} \left.
  - 3z \( 2z +\frac{1}{2} \) \log(1-z)
  + \( \frac{3}{2}\,z^2 +z+\frac{3}{2} \) \log(z)
   -\frac{\pi^2}{6} z \(3z+1\)  \rq a 
  \nonumber\\[2mm]
  &&{} + \ord{a^{\frac{3}{2}}\log(a)}
\end{eqnarray}

\begin{eqnarray} 
  ^{\sss\CA\!}g_{\sss 2}^{\sss C_1}(z) &=& -z \, \APreg_{qg}(z) \, a \log^2(a)
 \nonumber\\[2mm]
  &&{}
+\lq  2 z \log \frac{(1-z)^2}{z} \,\APreg_{qg}(z) +\frac{z}{2} (1-z) 
\rq a \log (a)
 \nonumber\\[2mm]
 &&{} + \bigg[4z \( \log (1-z)\log (z) -\log^2(1-z)-  \frac{1}{2}\log ^2(z) \)
 \APreg_{qg}(z)
\nonumber\\[2mm]
  &&{} \hspace{7.5mm}
-\frac{z}{2} (1-z)  \log (1-z) +\frac{1}{2} z \left(9 z^2-10 z+5\right) \log (z)
\nonumber\\[2mm]
&&{} \hspace{7.5mm}
-\frac{\pi^2}{3} z\,\APreg_{qg}(z)  +z\( 3 z^2 -2 z-1 \)
\bigg] a
 + \ord{a^{\frac{3}{2}}\log(a)} \phantom{aaaaaaa}
\end{eqnarray}

\begin{eqnarray}
  ^{\sss\CA\!}{\cal J}^{\sss C_1}_{\sss 23} &=&
   -\frac{1}{6} \lq\lumn + \lumnn \rq a\log^3(a)
  - \lq \frac{3}{4}\,\lumn + \frac{5}{3}\,\lumnn \rq a\log^2(a) 
  \nonumber\\[2mm]
  &&{} - \lq \( \frac{\pi^2}{2} + \frac{7}{2} \) \lumn
  + \(  \frac{\pi^2}{2} + \frac{241}{36} \) \lumnn \rq a\log(a) 
  \nonumber\\[2mm]
  &&{} - \lq \(  \frac{\pi^2}{12}\,+ 12 \) \lumn 
  + \(  \frac{4}{3}\,\pi^2 + \frac{1391}{108} \) \lumnn \rq a + \ord{a^{\frac{3}{2}}\log(a)}
\phantom{aaaaaaa}
\end{eqnarray}

\subsection[$C_1^{qg}$: $\CF$ coefficient]
           {$\boldsymbol{C_1^{qg}}$: $\boldsymbol{\CF}$ coefficient}
\begin{eqnarray} 
  ^{\sss\CF\!}g_{\sss 0}^{\sss C_1}(z) &=& -\frac{1}{3}\,\APreg_{qg}(z)\log^3(a)
  +  \APreg_{qg}(z) \lq \log(1-z) + \frac{3}{4} \rq \log^2(a)
  \nonumber\\[2mm]
  &&{} - \APreg_{qg}(z) \lq \log^2(1-z)
  +\frac{3}{2} \log(1-z) + \frac{7}{2} \rq \log(a)
  \nonumber\\[2mm]
  &&{} +  \APreg_{qg}(z) \( \frac{2}{3}\log^3(1-z)
 - \log(1-z)\log(z) \log\frac{1-z}{z}
 - \frac{1}{3}\log^3(z) \)
  \nonumber\\[2mm]
  &&{} + \( -\frac{3}{2}\,z^2 + z + \frac{1}{2} \) \log^2(1-z)
  + \( -3z^2 + \frac{5}{2}\,z - \frac{1}{4} \) \log^2(z)
  \nonumber\\[2mm]
  &&{}
  + \(6z^2-5z+\frac{1}{2}\) \log(1-z)\log(z)
   + \( -\frac{51}{4}\,z^2 + \frac{23}{2}\,z -\frac{9}{4} \) \log(z)
  \nonumber\\[2mm]
  &&{} + \( \frac{79}{4}\,z^2 - \frac{37}{2}\,z + \frac{23}{4}  \) \log(1-z)
  - \frac{93}{8}\,z^2 + \frac{37}{4}\,z + \frac{19}{8} 
  \nonumber\\[2mm]
  &&{} + \frac{1}{2}\,z \, a\log^2(a)
  + \lq -z \log(1-z) 
  -\frac{93}{16}\,z -\frac{19}{8} \rq a\log(a)
  \nonumber\\[2mm]
  &&{} + \lq -\frac{1}{2}\,z\log^2(z) - z\( \frac{23}{16}\,z^2
  +\frac{37}{8}\,z + \frac{31}{16} \) \log(z) + z
  \log(1-z)\log(z) \right.
  \nonumber\\[2mm]
  &&{} \hspace{7.5mm} \left. +\, \( \frac{23}{16}\,z^3 + \frac{37}{8}\,z^2 +
  \frac{81}{8}\,z + \frac{51}{8} \) \log(1-z)
  + \frac{23}{16}\,z^2 + \frac{19}{8}\,z \rq a
\nonumber\\[2mm]
  && {} + \ord{a^{\frac{3}{2}}\log(a)}
\end{eqnarray}

\begin{eqnarray} 
  ^{\sss\CF\!}g_{\sss 1}^{\sss C_1}(z) &=&
 -\frac{z}{2} \(3z+1\) a\log^2(a)
  \nonumber\\[2mm]
  &&{} + \lq 
  z\(3z+1\)\log(1-z)+\, \frac{47}{16}\,z^2 +\frac{59}{16}\,z +\frac{19}{8} \rq a\log(a)
  \nonumber\\[2mm]
  &&{} + \biggl[ - z \(3z+1\) \(  \log(1-z)\log\frac{1-z}{z}
    +\frac{1}{2} \log^2(z)  \)
  \nonumber\\[2mm]
  &&{} \hspace{7.5mm} 
  + \( \frac{23}{16}\,z^4 + \frac{51}{16}\,z^3 - \frac{9}{4}\,z - \frac{51}{8} \) \log(1-z)
  \nonumber\\[2mm]
  &&{} \hspace{7.5mm} \left.
   - \( \frac{23}{16}\,z^4 + \frac{51}{16}\,z^3 + \frac{9}{16}\,z^2 - \frac{3}{16}\,z \) \log(z)
   + \frac{139}{32}\,\,z^3 -\frac{877}{32}\,z^2 - \frac{105}{32}\,z \rq a 
   \nonumber\\[2mm]
  &&{} + \ord{a^{\frac{3}{2}}\log(a)}
\end{eqnarray}

\begin{eqnarray} 
  ^{\sss\CF\!}g_{\sss 2}^{\sss C_1}(z) &=& -z \, \APreg_{qg}(z) \, a \log ^2(a)
\nonumber\\[2mm]
  &&{} +
\lq 2 z \,\APreg_{qg}(z)\log (1-z)+\frac{1}{4} z \left(19 z^2-18 z+5\right) \rq
a \log (a) 
\nonumber\\[2mm]
  &&{} +
\bigg[  z\, \Big( 2 \log (z) \log (1-z)  -2  \log ^2(1-z)-  \log ^2(z)\Big)\, \APreg_{qg}(z)
  \nonumber\\[2mm]
  &&{} \hspace{7.5mm} -\frac{1}{4} z \left(19 z^2-18 z+5\right)
  \log (z)
\nonumber\\[2mm]
  &&{} \hspace{7.5mm} +\frac{1}{32} z \left(93 z^3-888 z^2+878 z-307\right)
\bigg] a + \ord{a^{\frac{3}{2}}\log(a)}
\end{eqnarray}

\begin{eqnarray}
  ^{\sss\CF\!}{\cal J}^{\sss C_1}_{\sss 23} &=&
  -\frac{1}{3} \lq \lumn + \lumnn \rq a\log^3(a)
   -\frac{1}{2} \lq \lumn +\frac{11}{6}\,\lumnn \rq a\log^2(a) 
  \nonumber\\[2mm]
  &&{} - \lq  \frac{47}{32} \,  \lumn 
   + \frac{341}{72} \, \lumnn \rq a\log(a) 
  \nonumber\\[2mm]
  &&{} + \lq \( - \frac{\pi^2}{3} + \frac{1013}{128} \) \lumn 
  - \frac{1}{36} \(  29\,\pi^2 +\frac{58691}{96} \) \lumnn \rq a + \ord{a^{\frac{3}{2}}\log(a)}
  \nonumber\\
\end{eqnarray}

\subsection[$C_2^{qg}$]{$\boldsymbol{C_2^{qg}}$}

\begin{eqnarray} 
  g_{\sss i}^{\sss C_2}(z) &=&  \frac{1}{3} \log\frac{\muf^2}{Q^2}\;
    g_{\sss i}^{\sss A}(z) \hspace{1.5cm} i=0,1,2
  \\[2mm]
 {\cal J}^{\sss C_2}_{\sss 23} &=& \frac{1}{3}  \log\frac{\muf^2}{Q^2}\; {\cal J}^{\sss A}_{\sss 23}
\end{eqnarray}

\providecommand{\href}[2]{#2}\begingroup\raggedright\endgroup

\end{document}